\documentclass[12pt,a4paper]{article}
\usepackage{epsfig}
\begin{document}
\title{Flavor changing $Z$ decay $Z\rightarrow b\overline{s}(\overline{b}s)$
in topcolor-assisted technicolor models  }

\author{Chongxing Yue$^{a }$,  Hong Li$^{b}$, Hongjie Zong$^{b}$  \\
%\address{
{\small $^{a}$Physics Department, Liaoning Normal University,
Dalian 116029, China
\thanks{E-mail:cxyue@lnnu.edu.cn}}\\
{\small $^{b}$ College of Physics and Information Engineering,}\\
\small{ Henan Normal University, Henan 453002, China}
 \\}
\date{\today}

\maketitle

\begin{abstract}
 In the context of topcolor-assisted technicolor (TC2) models, we examine
 the flavor changing (FC) $Z$ decay $Z\rightarrow b\overline{s}(\overline{b}s)$
 and calculate the contributions of the new particles predicted by TC2 models
 to the branching ratio Br($Z\rightarrow b\overline{s}+\overline{b}s$). We
 find that the contributions mainly come from the top-pions. In most of the
 parameter space, the Br($Z\rightarrow b\overline{s}+\overline{b}s$) can reach
 $10^{-5}$, which may be detected in near future experiments such as Giga-Z version
 of the TESLA. Thus, the FC $Z$ decay $Z\rightarrow b\overline{s}(\overline{b}s)$
 can be used to test TC2 models.
\end{abstract}

{\bf PACS number(s)}: 13.38.Dg, 12.15.Lk, 12.60.Nz, 14.80.Mz

\newpage

\vspace{.5cm}
\noindent{\bf I. Introduction}

~~It is widely believed that, in the standard model (SM), the
flavor-changing neutral currents (FCNC's) are absent at tree-level
and at one-loop level they are GIM suppressed. The SM results of
the rare decays, which are induced by FCNC, are very small and can
not be detected in the current or future experiments. The rare top
decays [1] and rare $Z$ decays [2] are two classes of such
examples. Thus, rare decays provide a very sensitive probe of new
physics beyond the SM. Detection of rare decays at visible levels
by any of the future colliders would be instant evidence of new
physics. Searching for rare decays is one of the major goals of
the next generation of high energy collider experiments.

Among rare decays the flavor-changing (FC) $Z$ decay $Z\rightarrow
q_{i}q_{j}$, which $q_{i}$ and $q_{j}$ are fermions of different
flavors, is one interest subject. Experimentally, high energy
$e^{+}e^{-}$ colliders can be used as $Z$ factory, providing an
opportunity to examine the decay properties of the neutral
electroweak gauge boson $Z$ in detail. The improved experimental
measurement at present stimulates the studies of these decays. For
example, with the Giga-$Z$ option of the TESLA linear collider
project, one may expect the production of about $10^{9}$ $Z$
bosons at resonance [3]. This huge rate allows one to study a
number of problems with unprecedented precision. Among them is the
search for the FC $Z$ decays.

We know that the dominate mode of the FC $Z$ decays is
$Z\rightarrow b\overline{s} (\overline{b}s)$. The branching ratio
Br($Z\rightarrow b\overline{s} +\overline{b}s$) in the SM has been
calculated in Ref.[2] and a lot of theoretical studies involving
the FC $Z$ decay $Z\rightarrow b\overline{s} (\overline{b}s)$have
been given within specific popular models beyond the SM. For
instance, the branching ratio Br($Z\rightarrow
b\overline{s}+\overline{b} s$)has been calculated in the two Higgs
doublets models (2HDM's)[4], in super-symmetry (SUSY)[5], SUSY
with $R$-parity violation [6], and in other beyond standard models
[7]. Recently, Ref.[8] has studied the FC $Z$ decay $Z\rightarrow
b\overline{s}(\overline{b}s)$ in the context of 2HDM's and SUSY
with flavor violation. They find that, within the SUSY scenarios
for flavor violation, the branching ratio Br($Z\rightarrow
b\overline{s}+\overline{b}s$) can reach $10^{-6}$ for large
$\tan\beta$ values.

The FC $Z$ decays may be useful in searching for new physics
beyond the SM at the TESLA collider or any other future colliders,
which are designed to run on the $Z$-pole with high luminosities.
With advances in technology, i.e., improved b-tagging
efficiencies, the FC $Z$ decay $Z\rightarrow b\overline{s}
(\overline{b}s)$, which is the easiest to detect among the FC
hadronic $Z$ decays, may be accessible to a Giga-$Z$ option even
for branching ratios as small as Br($Z\rightarrow
b\overline{s}+\overline{b}s$)$\sim 10^{-7}-10^{-6}$[8]. Thus, it
is very interesting to study this decay in various models beyond
the SM. The aim of this paper is to point out that, in the context
of topcolor-assisted technicolor (TC2) models [9], the branching
ratio Br($Z\rightarrow b\overline{s}+\overline{b}s$) also can be
significantly enhanced, which can reach the detectability
threshold for near future experiments such as Giga-Z version
 of the TESLA.

To completely avoid the problems arising from the elementary Higgs field in
the SM, various kinds of dynamical electroweak symmetry breaking (EWSB)
models have been proposed, and among which the topcolor scenario is
attractive because it explains the large top quark mass and provides possible
dynamics of EWSB. TC2 models [9], flavor-universal TC2 models [10] and top
see-saw models [11] are three of such examples. These kinds of models
generally predict the existence of colored gauge bosons (top-gluons, colorons)
, color-singlet gauge boson $Z^{\prime}$ and Pseudo-Goldstone bosons (technipions,
 top-pions). These new particles are most directly related to EWSB and
generation of fermion masses. Thus, studying the effects of these new particles
 in various processes would provide crucial information for EWSB
and fermion flavor physics as well. In this paper, we will
calculate the contributions of these new particles to the
branching ratio Br($Z\rightarrow b\overline{s}+\overline{b}s$). We
find that the contributions of  gauge bosons $B^{A}_{\mu}$ and
$Z^{\prime}$ to the Br($Z\rightarrow b\overline{s}+\overline{b}s$)
are small. The largest value is only $10^{-8}$. The main
contributions to the FC $Z$ decay $Z\rightarrow
b\overline{s}(\overline{b}s)$ come from the top-pions via the FC
scalar couplings. In most of the parameter space, the branching
ratio Br($Z\rightarrow b\overline{s}+\overline{b}s$) varies in the
range of $3.9\times10^{-5}\sim3.8\times10^{-6}$.

The paper is organized as follows: In section 2 we discuss the
contributions of Pseudo-Goldstone bosons (PGB's) to
Br($Z\rightarrow b\overline{s}+ \overline{b}s$). The effects of
topcolor gauge bosons on Br($Z\rightarrow
b\overline{s}+\overline{b}s$) are studied in section 3.
Discussions and conclusions are given in section 4.

\vspace{0.5cm}
\noindent{\bf II The contributions of PGB's}

In TC2 models [9], the TC interactions play a main role in breaking the
electroweak gauge symmetry. The ETC interactions give rise to the masses of
the ordinary fermions including a very small portion of the top quark mass,
namely $\varepsilon m_{t}$ with a model dependent parameter $\varepsilon\ll 1$.
The main part of the top quark mass is dynamically generated by topcolor
interactions at a scale of order $1TeV$, which also make small contributions
to EWSB. Thus, for TC2 models, there is the following relation:
\begin{equation}
v_{\pi}^{2}+F_{t}^{2}=v_{w}^{2} ,
\end{equation}
where $v_{\pi}$ represents the contributions of TC interactions to
EWSB, $v_{w}=v/\sqrt{2}=174GeV$, and  $F_{t}\approx50GeV$ is the
top-pion decay constant, which can be estimated from  the
Pagels-Stokar formula. This means that the associated top-pions
$\pi_{t}^{\pm,0}$ are not the longitudinal bosons $W$ and $Z$, but
are separately physically  observable objects. The presence of
physics top-pions in the low-energy spectrum is an inevitable
feature of topcolor scenario that purports to avoid fine-tuning
[12].

The flavor-diagonal couplings of top-pions to quarks can be written as
[9,12,13]:
\begin{eqnarray}
&&\frac{m_{t}(1-\varepsilon)}{\sqrt{2}F_{t}}
\frac{\sqrt{v_{w}^{2}-
F_{t}^{2}}}{v_{w}}[i\bar{t}\gamma^{5}t\pi_{t}^{0}+\sqrt{2}\bar{t}_{R}b_{L}
\pi_{t}^{+}+\sqrt{2}\bar{b}_{L}t_{R}\pi_{t}^{-}] \nonumber \\
&&\hspace{15mm}+\frac{m_{b}^{*}}{\sqrt{2}F_{t}}[i\bar{b}\gamma^{5}b\pi_{t}^{0}+\sqrt{2}
\bar{t}_{L}b_{R}\pi_{t}^{+}+\sqrt{2}\bar{b}_{R}t_{L}\pi_{t}^{-}],
\end{eqnarray}
where the factor $\frac{\sqrt{v_{w}^{2}-F_{t}^{2}}}{v_{w}}$
reflects the effect of the mixing between top-pions and the
Goldstone bosons of EWSB. From Eq.(2) we can see that the
couplings of top-pions $\pi_{t} ^{\pm}$ to the right-handed
b-quark ($b_{R}$) are very small, which are proportional to
$\frac{m_{b}^{*}}{\sqrt{2}F_{t}}(m_{b}^{*}\leq m_{b}\ll v_{w})$.
So, in our following calculation, we will ignore the couplings of
$\pi_{t}^{\pm}$ to $b_{R}$.

For TC2 models, the underlying interactions, topcolor interactions, are
non-universal and therefore do not posses GIM mechanism. When one writes the
 non-universal interactions in the quark mass eigen-basis, it can induce the
tree-level FC couplings. The FC couplings of top-pions to quarks can be
written as [14,15]:
\begin{eqnarray}
&&\frac{m_{t}}{\sqrt{2}F_{t}}\frac{\sqrt{v_{w}^{2}-
F_{t}^{2}}}{v_{w}}[iK_{UR}^{tc}K_{UL}^{tt^{*}}\bar{t}_{L}c_{R}\pi_{t}^{0}
+\sqrt{2}K_{UR}^{tc^{*}}K_{DL}^{bb}\bar{c}_{R}b_{L}\pi_{t}^{+}
+\sqrt{2}K_{UR}^{tc}K_{DL}^{bb^{*}}\bar{b}_{L}c_{R}\pi_{t}^{-}\nonumber \\
&&\hspace{15mm}+\sqrt{2}K_{UR}^{tc^{*}}K_{DL}^{ss}\bar{t}_{R}s_{L}\pi_{t}^{+}
+\sqrt{2}K_{UR}^{tc}K_{DL}^{ss^{*}}\bar{s}_{L}t_{R}\pi_{t}^{-}] ,
\end{eqnarray}
where $K_{UL(R)}$ and $K_{DL(R)}$ are rotation matrices that
diagonalize the up-quark  and down-quark mass matrices $M_{U}$ and
$M_{D}$, i.e., $K_{UL}^{+} M_{U}K_{UR}=M_{U}^{dia}$ and
$K_{DL}^{+}M_{D}K_{DR}=M_{D}^{dia}$, for which the CKM matrix is
defined as $V=K_{UL}^{+}K_{DL}$. To yield a realistic form of the
CKM matrix $V$, it has been shown [14] that the values of the
coupling parameters can be taken as:
\begin{equation}
K_{UL}^{tt}\approx K_{DL}^{bb} \approx K_{DL}^{ss}\approx1,
\hspace{10mm} K_{UR}^{tc}\leq\sqrt{2\varepsilon-\varepsilon^{2}} .
\end{equation}
In the following calculation, we will take $K_{UR}^{tc}=\sqrt{2\varepsilon-
\varepsilon^{2}}$ and take $\varepsilon$ as a free parameter, which is
assumed to be in the range of $0.03-0.1$ [9].

From Eq.(2) and Eq.(3), we can see that the FC $Z$ decay $Z\rightarrow
b\overline{s}+\overline{b}s$ can be induced through charged top-pion loops in
TC2 models. The relevant Feynman diagrams for the contributions of the
$\pi_{t}^{+}$ to $Z\rightarrow b\overline{s}$ are shown in Fig.1.

\begin{figure}[htb]
\vspace{0.5cm}
\begin{center}
\epsfig{file=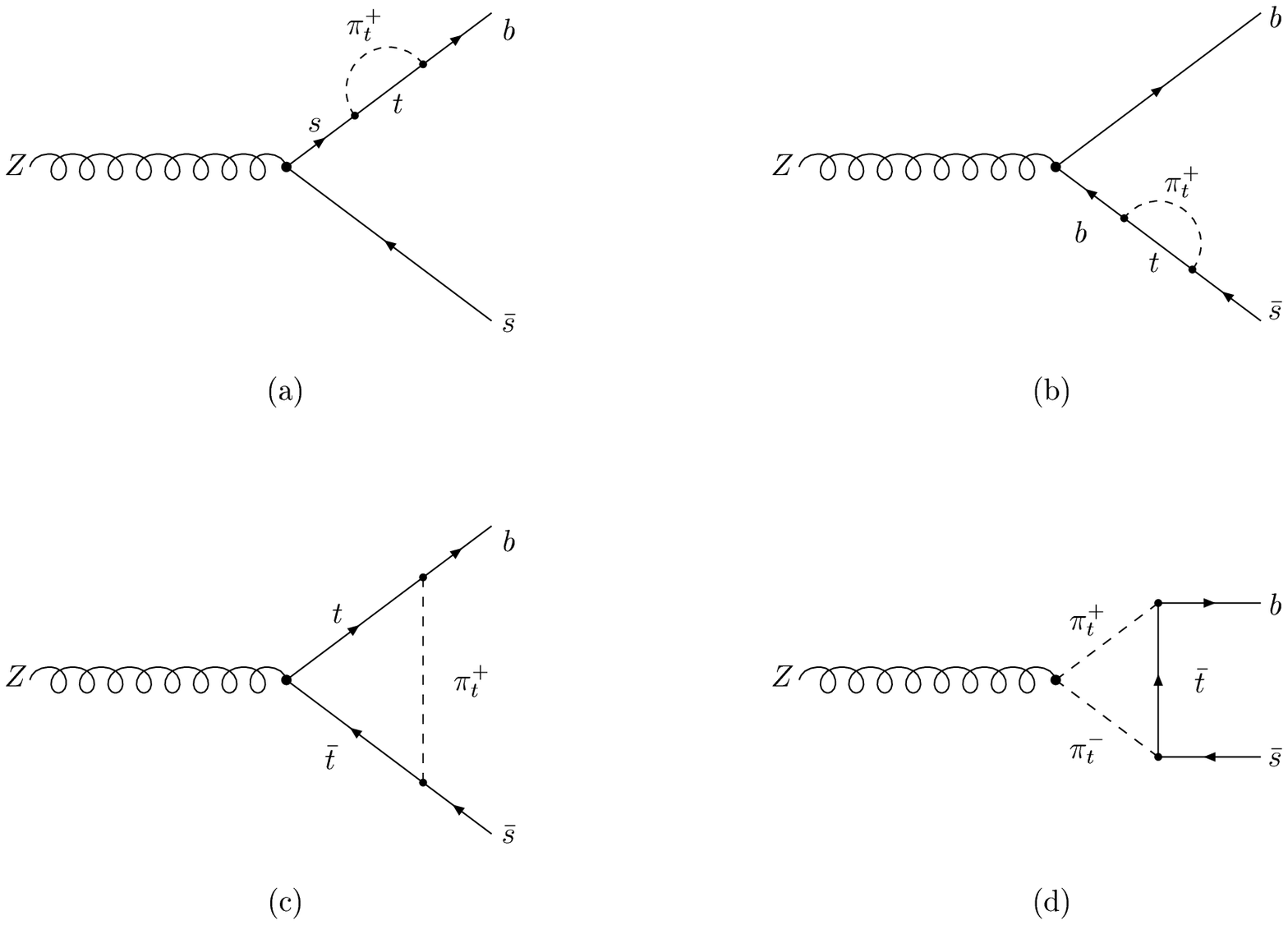,width=400pt,height=520pt} \vspace*{-8.5cm}
 \caption{Feynman diagrams that contribute to the flavor-changing (FC) $Z$
         decay $Z\rightarrow b\overline{s} $, due to charged
         top-pions $\pi_{t}^{\pm}$ exchange.}
 \label{ee}
\end{center}
\end{figure}

After a straightforward calculation, we can give the effective coupling
vertices $Z b\overline{s}$ in the on-shell renormalization scheme:
\begin{equation}
g_{L,\pi_{t}}^{bs}=\frac{m_{t}^{2}(1-\varepsilon)}{32\pi^{2}F_{t}^{2}}(1-
\frac{F_{t}^{2}}{v_{w}^{2}})\frac{e}{S_{W}C_{W}}K_{UR}^{tc}A,
\hspace{10mm}g_{R,\pi_{t}}^{bs} \approx 0,
\end{equation}
with
\begin{eqnarray}
A&=&\frac{\frac{2}{3}S_{W}^{2}-1}{m_{b}^{2}-m_{s}^{2}}
  [m_{b}^{2}B_{1}(m_{b}, m_{t}, m_{\pi_{t}})
  +m_{b}^{2}B_{0}(m_{b}, m_{t}, m_{\pi_{t}})\nonumber\\
&&+m_{s}^{2}B_{0}(m_{b}, m_{t}, m_{\pi_{t}})
  -m_{s}^{2}B_{0}(m_{s}, m_{t}, m_{\pi_{t}})\nonumber\\
&&-m_{s}^{2}B_{1}(m_{s}, m_{t}, m_{\pi_{t}})
  -m_{b}^{2}B_{0}(m_{s}, m_{t}, m_{\pi_{t}})]\nonumber\\
&&-(1-\frac{4}{3}S_{W}^{2})m_{t}^{2}C_{0}-\frac{4}{3}
S_{W}^{2}[B_{0}(m_{Z}, m_{t}, m_{t})-2C_{24}\nonumber\\
&&+m_{\pi_{t}}^{2}C_{0}]-2(1-2S_{W}^{2})C_{24}^{*},
\end{eqnarray}
where $S_{W}=\sin\theta_{W}$ and $C_{W}=\cos\theta_{W}$ which
$\theta_{W}$ is the Weinberg angle. $B_{i}$, $C_{0}$, and $C_{ij}$
are the standard two-point and three-point Feynman integrals [16].
$C_{0}=C_{0}(m_{s},m_{Z},m_{\pi_{t}},m_{t},m_{t})$, $C_{24}=C_{ij}
(m_{s},m_{Z},m_{\pi_{t}},m_{t},m_{t})$ and
$C_{24}^{*}=C_{ij}(m_{s}, m_{Z},m_{t},m_{\pi_{t}},m_{\pi_{t}}).$
The total decay width of the FC $Z$ decay $Z\rightarrow
b\overline{s}+ \overline{b}s$ can be written as:
%\begin{equation}
\begin{eqnarray}
\Gamma(Z\rightarrow
b\overline{s}+\overline{b}s)=\frac{\alpha_{e}m_{t}^{4}
m_{Z}(1-\varepsilon)^{2}}{96\pi^{4}F_{t}^{4}}\frac{1}{(4S_{W}C_{W})^{2}}
(1-\frac{F_{t}^{2}}{v_{w}^{2}})^{2} (K_{UR}^{tc})^{2}A^{2}.
%\end{equation}
\end{eqnarray}
In above equation, we have taken the approximations such as
$m_{Z}^{2}-m_{b}^{2}-m_{s}^{2}\approx m_{Z}^{2}$.

In Fig.2 we plot Br($Z\rightarrow b\overline{s}+\overline{b}s$) as
a function of  $m_{\pi_{t}}$ for three values of the parameter
$\varepsilon$: $\varepsilon= 0.03, 0.05$ and $0.08$. In our
calculation, we have taken: $\Gamma_{Z}=2.49GeV$, $m_{t}=175GeV$,
$m_{b}=4.8GeV$, $m_{Z}=91.18GeV$, $m_{s}= 0.15GeV$,
$\alpha_{e}=\frac{1}{128}$, and $S_{W}^{2}=0.2322$ [17]. From
Fig.2 we can see that Br($Z\rightarrow
b\overline{s}+\overline{b}s$) increases with the parameter
$\varepsilon$ increasing. On the other hand, the branching ratio
is sensitive to the top-pion mass $m_{\pi_{t}}$ and strongly
suppressed by large $m_{\pi_{t}}$. For $200GeV\leq
m_{\pi_{t}}\leq450GeV$ and $0.03\leq\varepsilon\leq0.08$,
Br($Z\rightarrow b\overline{s}+\overline{b}s$) varies in the range
of $3.9\times10^{-5}\sim3.8\times10^{-6}$.

To solve the phenomenological difficulties of the traditional TC
models [13,18], TC2 models [9] were proposed by combing
technicolor interactions with the topcolor interactions for the
third generation quark at the scale about $1TeV$. Thus, TC2 models
predict number of technipions in the technicolor sector. These new
particles also have contributions to the FC $Z$ decay
$Z\rightarrow b\overline{s}$ via the coupling $\pi^{+}u_{i}d_{j}$.
However, in TC2 models, the technipion-top-bottom coupling is
proportional to $\frac{\varepsilon m_{t}}{F_{\pi}}$ and the
technipion contributions to Br($Z\rightarrow
b\overline{s}+\overline{b}s$) are proportional to
$(\frac{\varepsilon m_{t}}{F_{\pi}})^{2}$. Furthermore, the
coupling $\pi^{+}\bar{t}s$ is proportional to CKM matrix element
$V_{ts}$, which is smaller than $K_{UR}^{tc}$. Thus, the
contributions of technipions to Br($Z\rightarrow b\overline{s}+
\overline{b}s$) are very small, which can be ignored.

\begin{figure}[htb]
\vspace*{0cm}
\begin{center}
\vspace*{0cm}
 \epsfig{file=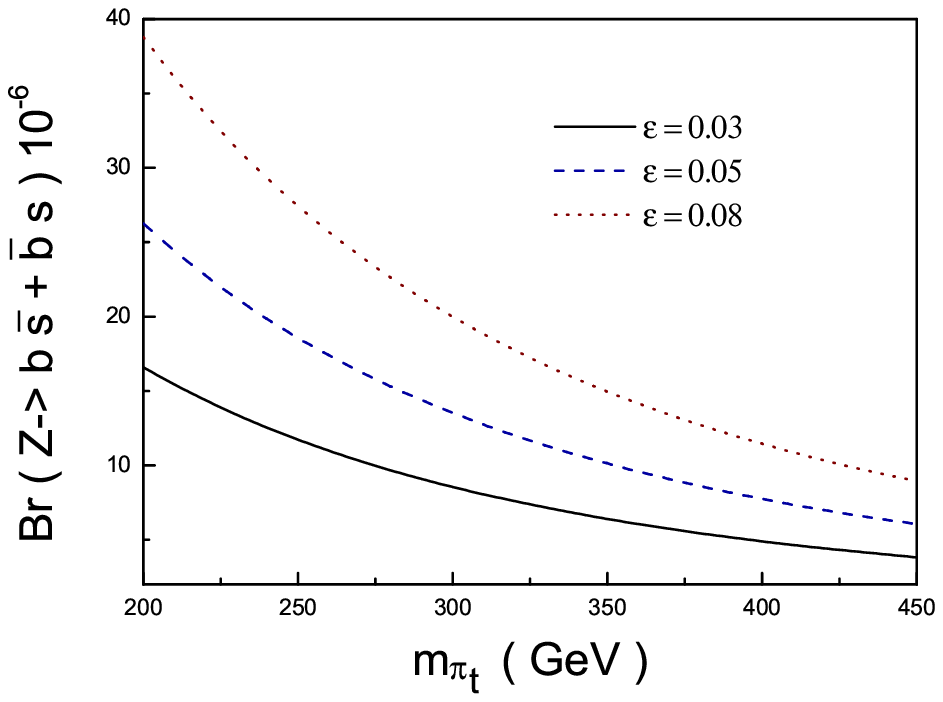,width=450pt,height=300pt}
\vspace*{-1cm}
\caption{The branching ratio Br($Z\rightarrow
b\overline{s}+\overline{b}s$)
         contributed by top-pions $\pi_{t}^{\pm}$ as a function of
         $m_{\pi_{t}}$ for the parameter $\varepsilon=0.03$(solid line),
         $\varepsilon=0.05$ (dash line), and $\varepsilon=0.08$(dotted line).}
\label{fig.2}
\end{center}
\end{figure}

\vspace{.5cm}
\noindent{\bf III. The contributions of topcolor gauge bosons}

The key feature of TC2 models [9] is that a large part of the top
quark mass is dynamically generated by topcolor interactions at a
scale of order $1TeV$, which is flavor non-universal. To ensure
that the top quark condenses and receives a large mass while the
bottom quark does not, a non-universal extended hyper-charge group
$U(1)$ is often invoked, so that the topcolor gauge group is
usually taken to be a strong coupled $SU(3)\times U(1)$. At the
$\Lambda\sim 1TeV$, the dynamics of a general TC2 model involves
the following structure [13,15]:
\begin{eqnarray}
SU(3)_{1}\times SU(3)_{2}\times U(1)_{Y_{1}}\times U(1)_{Y_{2}}\times SU(2)_{L}
\rightarrow  SU(3)_{QCD}\times U(1)_{EM},
\end{eqnarray}
where $ SU(3)_{1}\times U(1)_{Y_{1}} ( SU(3)_{2}\times U(1)_{Y_{2}})$
 generally couples preferentially to the third (first and second) generations.
The $U(1)_{Y_{i}}$ are just strongly rescaled versions of
electroweak $U(1)_{Y}$. This breaking scenario gives rise to the
topcolor gauge bosons including the color-octet colorons
$B_{\mu}^{A}$ and color-singlet extra $U(1)$ gauge boson
$Z^{\prime}$. As a result, these new massive gauge bosons
$B_{\mu}^{A}$ and $Z^{\prime}$ couple predominantly to the third
generation quarks and the third generation fermions, respectively.
The flavor-diagonal couplings of these new gauge bosons to quarks,
which are related to the FC $Z$ decay $Z\rightarrow b\overline{s}
(\overline{b}s)$, can be written as:
\begin{eqnarray}
{\cal L} _{B}^{FD}&=&\frac{1}{2}g_{3}B_{\mu}^{A}[\cot\theta\bar{b}
\gamma^{\mu}\lambda^{a}b-\tan\theta\bar{s}\gamma^{\mu}\lambda^{a}s],\\
{\cal L}_{Z^{\prime}}^{FD}&=&\frac{1}{6}g_{1}\cot\theta^{\prime}
Z_{\mu}^{\prime}(\bar{b}_{L}\gamma^{\mu}b_{L}-2\bar{b}_{R}\gamma^{\mu}b_{R}) \nonumber\\
&&-\frac{1}{6}g_{1}\tan\theta^{\prime}
  Z_{\mu}^{\prime}(\bar{s}_{L}\gamma^{\mu}s_{L}-2\bar{s}_{R}\gamma^{\mu}s_{R}),
\end{eqnarray}
with
\begin{equation}
k_{3}=\frac{g_{3}^{2}\cot^{2}\theta}{4\pi},\hspace{10mm}k_{1}=
\frac{g_{1}^{2}\cot^{2}\theta^{\prime}}{4\pi}.
\end{equation}
Where $g_{3}(g_{1})$ is the QCD ($U(1)_{Y}$) coupling constant at
$\Lambda_{TC}$, $\theta $ and $\theta'$ are mixing angles. To
select the top quark direction for condensation and not form a
$b\bar{b}$ condensation, there must be $\cot\theta\gg 1$ and
$\cot\theta^{\prime}\gg 1$. To obtain proper vacuum tilting, the
coupling constants $k_{3}$ and $k_{1}$ should satisfy certain
constraint. There is a region of $k_{3}$ and $k_{1}$, i.e.,
$k_{3}=2$, $k_{1}\leq 1$, satisfying requirement of vacuum tilting
and the constraints from $Z$-pole physics and $U(1)$ triviality
shown in Refs.[10,15]. We shall take $k_{3}=2$ and $k_{1}=1$ in
the following calculation.

Similar to the top-pions, when one writes the non-universal
interactions in the quark mass eigen-basis, these interactions can
result in the FCNC vertices of the new gauge bosons . The FC
couplings of the gauge bosons $B_{\mu}^{A}$ and $Z^{\prime}$ to
quarks, which are related to our calculation, can be written as :
\begin{eqnarray}
{\cal L} _{B}^{FC}&=&\frac{1}{2}g_{3}K_{B}^{bs}B_{\mu}^{A}\bar{b}
\gamma^{\mu}\lambda^{a}s,\\
{\cal
L}_{Z^{\prime}}^{FC}&=&-\frac{1}{6}g_{1}K_{Z}^{bs}Z_{\mu}^{\prime}
(\bar{b}_{L}\gamma^{\mu}s_{L}-2\bar{b}_{R}\gamma^{\mu}s_{R}),
\end{eqnarray}
where $K_{B}^{bs}$ and $K_{Z}^{bs}$ are the flavor fixing factors. In the
following estimation, we will assume $K_{B}^{bs}=K_{Z}^{bs}=K$.

The Feynman diagrams, which represent the new gauge boson exchange
contributions to the process $Z\rightarrow b\bar{s}$, are depicted
in Fig.3.
\begin{figure}[htb]

\begin{center}
\vspace*{-3.5cm} \epsfig{file=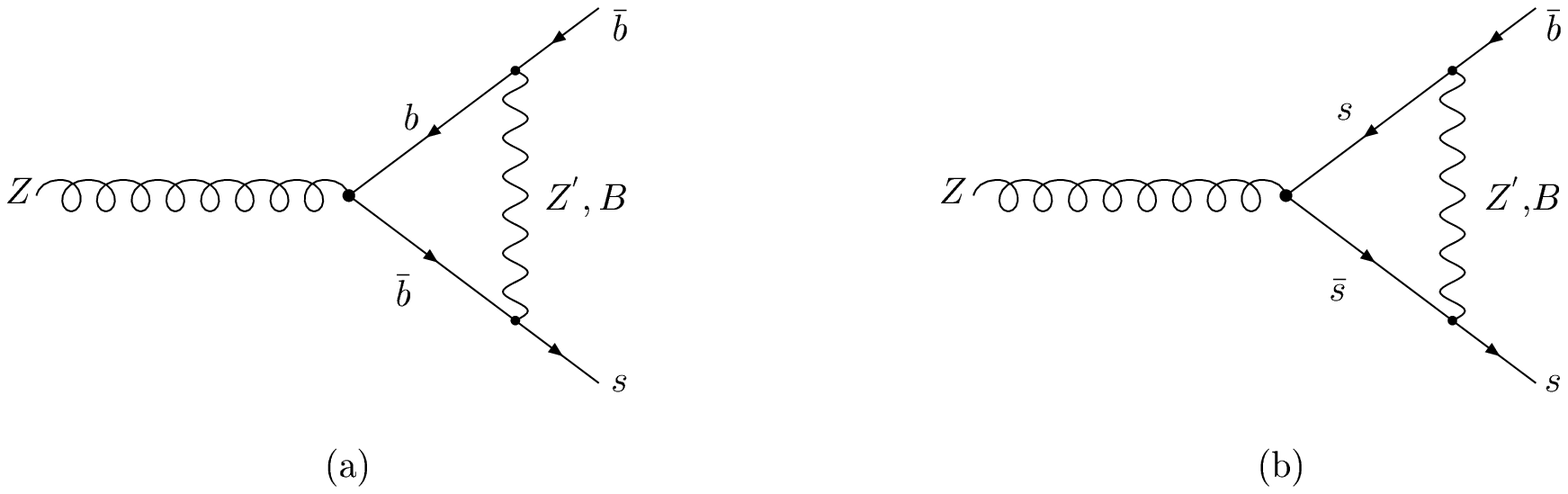,width=400pt,height=520pt}
\vspace*{-8.5cm}
 \caption{Feynman diagrams for the contributions of the topcolor gauge bosons
         $B_{\mu}^{A}$ and $Z^{\prime}$ to the FC $Z$ decay $Z\rightarrow
         b\overline{s} $.}
 \label{Fig3}
\end{center}
\end{figure}

 Similar to Ref.[19], we can calculate these diagrams straightforwardly. In our calculation,
 we have taken
$m_{b}\approx 0$, $m_{s}\approx 0$. It is easy to see that,
relative to the contributions of Fig.3(a), the contributions of
Fig.3(b) to $Z\rightarrow b\bar{s}$ are suppressed by the factors
$\tan^{2}\theta$ and $\tan^{2} \theta^{\prime}$, which correspond
gauge bosons $B_{\mu}^{A}$ and $Z^{\prime}$, respectively. Then
the effective couplings of $Zb\bar{s}$, which arise from the gauge
bosons $B_{\mu}^{A}$ and $Z^{\prime}$ can be written as:
%\begin{eqnarray}
\begin{equation}
g_{L,B}^{bs}=g_{R,B}^{bs}=\frac{2k_{3}\tan\theta K}{9\pi}g_{L}^{b}[\frac{m_{Z}^{2}}
{M_{B}^{2}}\ln\frac{M_{B}^{2}}{m_{Z}^{2}}],
\end{equation}
\begin{equation}
 g_{L,Z}^{bs}=\frac{k_{1}\tan\theta^{\prime} K}{54\pi}g_{L}^{b}[\frac{m_{Z}^{2}}{M_{Z}^{2}}
 \ln\frac{M_{Z}^{2}}{m_{Z}^{2}}],  \hspace{8mm}
g_{R,Z}^{bs}=\frac{2k_{1}\tan\theta^{\prime} K}{27\pi}g_{R}^{b}
[\frac{m_{Z}^{2}}{M_{Z}^{2}}\ln\frac{M_{Z}^{2}}{m_{Z}^{2}}],
\end{equation}
%\end{eqnarray}
with
\begin{equation}
g_{L}^{b}=\frac{e}{S_{W}C_{W}}(-\frac{1}{2}+\frac{1}{3}S_{W}^{2}),\hspace{10mm}
g_{R}^{b}=\frac{e}{S_{W}C_{W}}(\frac{1}{3}S_{W}^{2}).
\end{equation}

The limits on the masses of the topcolor gauge bosons
$B_{\mu}^{A}$ and $Z^{\prime}$ can be obtained via studying their
effects on various experimental observable [13]. For example,
Ref.[20] has shown that $B\bar{B}$ mixing provides stronger lower
bounds on the masses of $B_{\mu}^{A}$ and $Z^{\prime}$, one must
have $M_{B}>3.1TeV$ ($4.8TeV$) and $M_{Z}>6.8TeV$ ($9.6TeV$) if
ETC does (does not) contribute to the CP-violation parameter
$\epsilon$. Recently, Ref.[21] restudy the bound placed by the
electroweak measurement data on the extra $U(1)$ gauge boson
$Z^{\prime}$. They find that $Z^{\prime}$ predicted by TC2 models
must be heavier than about $1TeV$. As estimation the contributions
of the topcolor gauge bosons to the FC $Z$ decay $Z\rightarrow
b\overline{s}(\overline{b}s)$, we take $M_{Z}=M_{B}=M$ and take
the mixing factor $K_{Z}^{bs}=K_{B}^{bs}=K$ as free parameters in
this paper.

Using Eq.(14)-(16), we can give the values of the branching ratio
Br ($Z\rightarrow b\overline{s}+\overline{b}s$) arised from the
topcolor gauge boson exchange. Our numerical results are shown in
Fig.4 and Fig.5,
\begin{figure}[htb]
\vspace*{-0.5cm}
\begin{center}
\epsfig{file=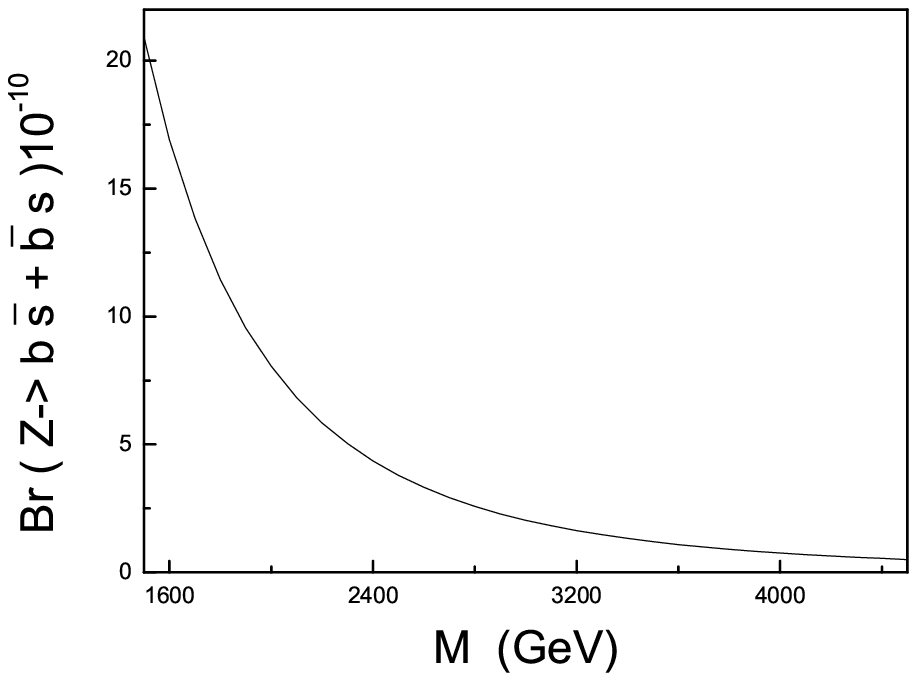,width=450pt,height=300pt}
  \end{center}
\vspace*{-1.8cm}
\caption{Br($Z\rightarrow b\overline{s}+\overline{b}s$) contributed by
         $B_{\mu}^{A}$ and $Z^{\prime}$ as a function of $M$ for
         $K=\lambda=0.22$.}
\label{fig.4}
 \vspace{-6cm}
 \end{figure}
\begin{figure}[htb]
\vspace*{0cm}
 \begin{center}
\epsfig{file=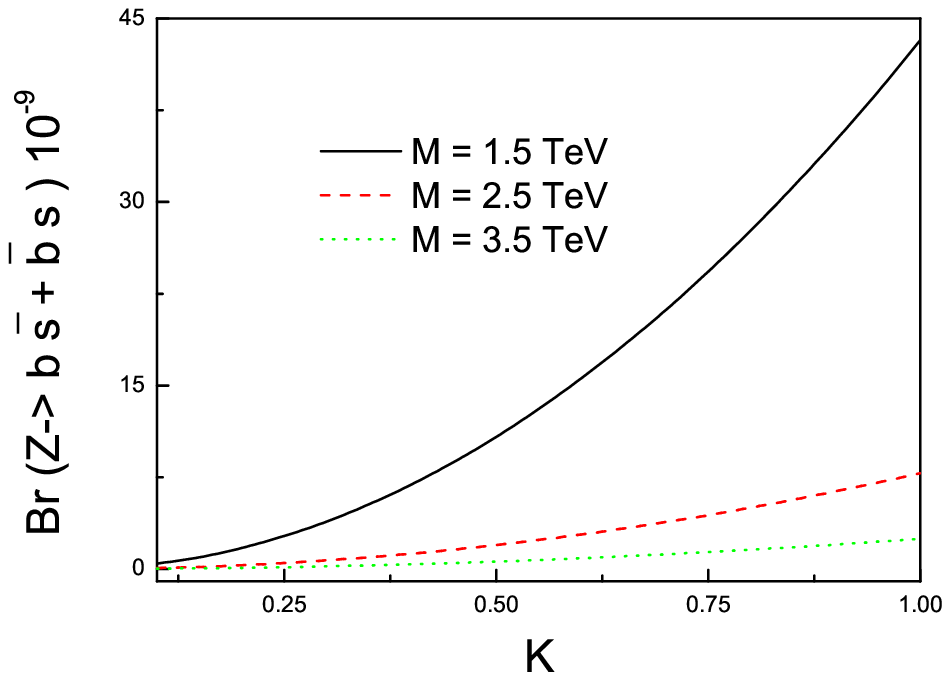,width=450pt,height=300pt}
\end{center}
\vspace{-1cm}
\caption{Br($Z\rightarrow b\overline{s}+\overline{b}s$) contributed by
         $B_{\mu}^{A}$ and $Z^{\prime}$ as a function of the flavor mixing
         factor $K$ for $M=1.5TeV$ (solid line), $M=2.5TeV$ (dash line),
         and $M=3.5TeV$ (dotted line).}
\label{fig.5}
  \vspace*{0cm}
\end{figure}
\vspace{5.5cm}
in which we plot Br($Z\rightarrow
b\overline{s}+\overline{b}s$) as a function of the mass $M$ for
$K=\lambda$($\lambda=0.22$ is the Wolfenstein parameter [22]) and as a function
of the mixing factor $K$ for $M=1.5TeV$, $2.5TeV$ and $3.5TeV$, respectively.
We can see from Fig.4 and Fig.5 that the branching ratio
Br($Z\rightarrow b\overline{s}+\overline{b}s$) decreases with $M$ increasing
and $K$ decreasing. In all of the parameter space, the Br($Z\rightarrow
b\overline{s}+\overline{b}s$) is smaller than $10^{-7}$. For $M=1.5TeV$, the value of
Br($Z\rightarrow b\overline{s}+\overline{b}s$) can reach $2.1\times10^{-9}
(4.3\times10^{-8})$ for $K=0.22(1)$.

The flavor-universal TC2 models [10] also predict the presence of
color-octet gauge bosons (topgluons) and color-singlet gauge boson
$Z^{\prime}$. However, the topcolor interactions are
flavor-universal and all quarks carry the same $SU(3)$ charge, the
topgluons couple with equal strength to all quarks. As a result,
topgluons can not cause tree-level FCNC's and have no
contributions to the FC $Z$ decay $Z\rightarrow
b\overline{s}(\overline{b}s)$. To ensure that top quark condenses
and receives a large mass while the b-quark does not, a
non-universal extended hyper-charge group $U(1)$ is involved in
the flavor-universal TC2 models. Thus, the $Z^{\prime}$ predicted
by these models treats the third generation fermions differently
than those in the first and second generations and can cause the
tree-level FCNC's. So the gauge boson $Z^{\prime}$ can give
contributions to Br($Z\rightarrow b\overline{s}+ \overline{b}s$).
The numerical results are plotted in Fig.6 for the values of the
$Z^{\prime}$ mass: $M=1.5TeV$, $2.5TeV$ and $3.5TeV$. We can see
from Fig.6 that the  contributions of $Z^{\prime}$ to
Br($Z\rightarrow b\overline{s}+\overline{b}s$) are very small. The
largest allowed value for the Br($Z\rightarrow
b\overline{s}+\overline{b}s$) is $\sim10^{-11}$. For example, the
value of the Br($Z\rightarrow b\overline{s}+\overline{b}s$) is
only $9.5\times10^{-13}$ for $M_{Z}=1.5TeV$ and $K=\lambda=0.22$.
\begin{figure}[htb]
\vspace*{0cm}
\begin{center}
\epsfig{file=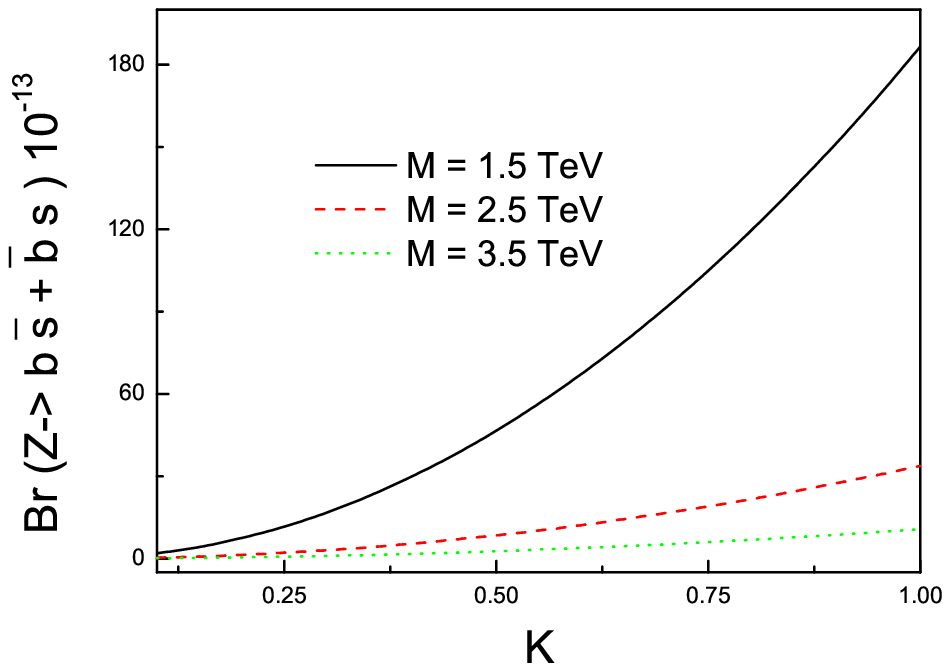,width=450pt,height=270pt}
\vspace{-1cm}
\end{center}

\caption{Br($Z\rightarrow b\overline{s}+\overline{b}s$) contributed by
         $Z^{\prime}$ as a function of the flavor mixing factor $K$ for
         $M=1.5TeV$ (solid line), $M=2.5TeV$ (dash line), and
         $M=3.5TeV$ (dotted line) in the flavor-universal TC2 models.}
\label{fig.6}
\vspace{0cm}
\end{figure}

The extra $U(1)$ gauge bosons $Z^{\prime}$ predicted by
non-commuting ETC model [23] and un-unified standard model [24],
which couples differently to fermions, also has contributions to
the FC $Z$ decay $Z\rightarrow b\overline{s} (\overline{b}s)$.
However, its contributions to Br($Z\rightarrow
b\overline{s}+\overline{b}s$) are also very small.

\vspace{.5cm} \noindent{\bf IV. Discussions and conclusions}

The top quark, with a mass of the order of the electroweak scale,
is singled out to play a key role in the dynamics of EWSB and
flavor symmetry breaking. There may be a common origin for EWSB
and top quark mass generation. The idea is first accomplished in
the top-condensation model [25]. However, this model can not fully
break the electroweak symmetry and also generate a large top quark
mass . To address this problem, various kinds of strong top
dynamical models have been proposed, including TC2 models [9],
flavor-universal TC2 models [10], top see-saw models [11], and the
top flavor seesaw models [26]. The common feature of such type of
models is that the topcolor interactions generate the main part of
top quark mass and also make small contributions to EWSB. EWSB is
mainly generated by TC interactions or other interactions. Then,
the presence of physical top-pions in the low-energy spectrum is
an inevitable feature of these models. The effects of top-pions on
low-energy observables are governed by its mass $m_{\pi_{t}}$,
while the large couplings of top-pions to quarks and to gauge
bosons are to large degree model-independent [12]. Thus, our
calculations about the contributions of top-pions to the FC $Z$
decay $Z\rightarrow b\overline{s}(\overline{b}s)$ also apply to
other models. Certainly, the relevant flavor mixing factor may has
different values for different models.

Another common feature of strong top dynamical models is that they
extended one or more of the SM $SU(N)$ gauge groups to an
$SU(N)\times SU(N)$ structure at energies well above the
electroweak scale [27]. All of these models propose that the gauge
groups should be flavor non-universal. For example, $SU(3)$ gauge
group is flavor non-universal in TC2 models and $U(1)$ gauge group
is flavor non-universal in the flavor-universal TC2 models and TC2
models. When the non-universal  interactions are written in the
mass eigen-states, the corresponding gauge bosons can induce the
tree-level flavor changing couplings. Then these new gauge bosons
may have significant contributions to some FCNC's processes. Our
numerical results show that the extra color-octet gauge boson
$B_{\mu}^{A}$ predicted by TC2 models and the extra $U(1)$
 gauge boson $Z^{\prime}$ predicted by TC2 models or
the flavor-universal TC2 models can indeed give contributions to
the FC $Z$ decay $Z\rightarrow b\overline{s}(\overline{b}s)$. With
reasonable values of the parameters, the branching ratio
Br($Z\rightarrow b\overline{s}+\overline{b}s$ ) can reach
$4.3\times 10^{-8}$ for TC2 dynamics. However, it is smaller than
the reach of a Giga-$Z$ $e^{+}e^{-}$ collider.

 To summarize, we have examined the FC $Z$ decay process
$Z\rightarrow b\overline{s}(\overline{b}s)$ in the framework of
TC2 models and calculated the contributions of the new particles
predicted by TC2 models to the branching ratio Br($Z\rightarrow
b\overline{s}+\overline{b}s$). We find that the contributions of
top-pions are larger than those of topcolor gauge bosons. In
reasonable parameter space of TC2 models, the value of
Br($Z\rightarrow b\overline{s}+\overline{b}s$) can be reach
$3.9\times 10^{-5}$, which may be detected by the Giga-$Z$ TESLA
colliders. Thus, a signal of $Z\rightarrow b\overline{s}
(\overline{b}s)$ in a Giga-$Z$ TESLA or any other colliders will
be consistent with the underlying mechanisms for EWSB and top
quark mass generation in topcolor scenario.

\vspace{.5cm}
\noindent{\bf Acknowledgments}

Chongxing Yue thanks the Abdus Salam International Centre for
Theoretical Physics (ICTP) for partial support. This work was
supported in part by the National Natural Science Foundation of
China (90203005) and Foundation of Henan Educational Committee.

%\newpage

\null

\end{document}